\documentclass[preprint,review,12pt]{elsarticle}
\usepackage{sectsty}

\usepackage[margin=2.5cm]{geometry}
\usepackage{setspace}

\usepackage{titlecaps}

\usepackage{titlesec}

\titleformat*{\section}{\MakeUppercase}
\titleformat*{\subsection}{\bfseries}
\titleformat*{\subsubsection}{\itshape}
\usepackage{float}
\usepackage{multirow}
\usepackage{textcomp}
\usepackage{graphicx}
\usepackage{amsmath}
\usepackage{longtable}
\usepackage{textcomp}
\usepackage{bm}
\usepackage{natbib}
\usepackage{epstopdf}
\journal{The CJCE Special Issue on CFD Symposium at WCCE-10}

\biboptions{numbers,sort&compress}
\biboptions{super}
\biboptions{square}

\begin{document}

\begin{frontmatter}
\title{CFD study on Taylor bubble characteristics in Carreau\textendash Yasuda shear thinning liquids}
\author{Somasekhara Goud Sontti}
\author{Arnab Atta\corref{cor1}}
\cortext[cor1]{Corresponding author. Tel.: +91 3222 283910}
\ead{arnab@che.iitkgp.ac.in}
\address{Multiscale Computational Fluid Dynamics (mCFD) Laboratory, Department of Chemical Engineering, Indian Institute of Technology Kharagpur, West Bengal 721302, India}

\begin{abstract}
\begin{spacing}{1}	
In the present study, Taylor bubble formation in two\textendash phase gas\textendash non-Newtonian Carreau liquid flowing through a confined co\textendash flow microchannel is investigated. Systematic analysis are carried out to explore the influences of rheological properties, inlet velocities, and surface tension on Taylor bubble length, shape, velocity and liquid film thickness. Aqueous solutions of carboxymethyl cellulose (CMC) with different mass concentrations are considered as the non\textendash Newtonian liquids to understand the fundamentals of flow behaviour. With increasing solution viscosity and liquid phase inlet velocity, Taylor bubble formation frequency and velocity increased, however, the bubble length was found to decrease. Velocity profiles inside the Taylor bubble and liquid slug were analyzed, and distinct velocity distributions were found for different CMC concentrations. Flow pattern maps are constructed based on inlet velocities for Carreau liquids in co\textendash flow microchannel. This study essentially provides useful guidelines in designing non-Newtonian microfluidic system for precise control and manipulation of Taylor bubbles.
\end{spacing}
\end{abstract}

\begin{keyword} 
Microchannel \sep Shear thinning liquid \sep Taylor bubble \sep Co\textendash flow, CFD
\end{keyword}

\end{frontmatter}

\section*{Introduction}

In recent years, studies on two\textendash phase flow in microchannels have attracted vast interest due to its wide range of applications in microfluidics, lab-on-a-chip devices, and microreactors.\citep{kreutzer-2005,liu2017m} Two\textendash phase flow patterns are generally classified as bubbly, Taylor, churn, annular, and stratified flow. Taylor bubble flow is one of the critical two\textendash phase flow patterns, where elongated bubbles are separated by liquid slug, and is characterized by a length larger than the equivalent diameter of the channel. Significant advantages of the Taylor bubble flow are its superior heat and mass transfer performance, which are distinguished by interfacial area and internal circulation.\citep{kreut-2005,gunther2006,abiev-2012,pan2014} These parameters are determined by the bubble length, near-wall liquid film thickness, and velocity field inside the Taylor bubble, as well as in liquid slug. Internal circulatory flow patterns enhance convective mass transport, and reduce axial dispersion.\citep{irandoust1992} Co\textendash flow configuration is one of the simplest microfluidic geometries that is typically utilized for Taylor bubble formation, where the dispersed and continuous phases flow in parallel to one another. To obtain the desired segmented flows, specific ranges of flow rates for both phases need to be controlled. Numerous studies have proposed various flow regimes for different flow condition in microchannels.\citep{ kawah-2002, chung-2004, liuuu-2005} Salman et al.~\citep{salman-2006} studied Taylor bubble formation in capillary tubes using co\textendash flow microchannel, and observed different mechanism of bubble formation and coalescence for low liquid flow rates. To identify the interface between two phases, different numerical methods are typically available namely, Front-Tracking (FT) \citep{van2006numerical}, Volume of Fluid (VOF) \citep{hirt-1981,van2005numerical}, Level Set (LS) \citep{suss-1994}, Phase Field \citep{sun-2007} and Lattice Boltzmann method (LBM) \citep{shan-1993l,shu2013direct}. Yu et al.~\citep{yu2007experiment} carried out experiments and LBM simulations of air--oil two-phase flow in cross and converging shaped microchannels to understand the bubble shape, size, and formation mechanism under different flow rates and mixer geometries. They identified bubbly and slug flow regimes depending on the Capillary number (Ca). Chen et al.~\citep{chen-2007} investigated Taylor bubble formation in a nozzle\textendash tube co\textendash flow configuration using LS method, and predicted liquid film thickness around the bubble, which was in fair agreement with the experimental observation by Bretherton~\citep{breth-1961}. Goel and Buwa\citep{goel-2008} analyzed bubble formation in circular capillaries using VOF method, and described the effect of various parameters, such as superficial velocities, capillary diameter, and wall contact angle. With the help of a VOF model, Gupta et al.~\citep{gupta-2009} critically analysed Taylor bubble flow in a circular microchannel, and provided guidelines to capture liquid film thickness around the Taylor bubble by imposing refined mesh in the vicinity of wall. Wang~\citep{wangg-2015} also applied VOF method to understand Taylor bubble flow for air\textendash water system in a tapered co\textendash flow geometry, where the effects of flow rates and nozzle injection length on bubble sizes and bubbling frequencies were explored.

\noindent From the literature, it is apparent that most of the efforts are devoted in understanding and modelling of the Taylor flow hydrodynamics and its characteristics associated with Newtonian liquids at the microscale. However, the fluid behaviour of non\textendash Newtonian fluids are known to be different from that of Newtonian fluid, as the viscosity depends on the shear rate.\citep{azaiez2007, govindarajan2014} In reality, many commercial chemicals (e.g., colloidal suspensions and polymer solutions) and bio\textendash logical samples (e.g., blood and DNA solutions) exhibit non\textendash Newtonian shear thinning and viscoelastic characteristics.\citep{nghe2011} Bubble formation and breakup mechanism in such liquids are complex due to distinctive attributes of non\textendash Newtonian liquids.\citep{frank2003} Consequently, the fundamental studies of two\textendash phase flow involving non\textendash Newtonian fluids are of paramount interest, and considerable attention have been devoted to gain insights into the effect of rheological properties on Taylor bubble formation in different microchannel configurations.\citep{hemminger2010,fu-2011,chen2013,labor-2015} Fu et al.~\citep{fu-2011} experimentally studied bubble formation in T\textendash junction microchannels using different concentrations of polyacrylamide (PAAm) solutions, and observed various flow patterns by varying gas and liquid flow rates. It is evident from their results that the bubble size increases non\textendash linearly with the gas-liquid flow rate ratios, and decreases with the concentration of PAAm solutions. Chen et al.~\citep{chen2013} developed a three\textendash dimensional numerical model to understand bubble formation in a T\textendash junction microchannel for Newtonian and non\textendash Newtonian liquids using VOF method. Initially, their numerical model was verified for Newtonian liquids with in\textendash house experimental visualization, and then the model was extended for power\textendash law and Bingham liquids. Laborie et al.~\citep{labor-2015} illustrated the effect of yield stress fluids on bubble formation in T\textendash junction, and flow\textendash focusing microchannels. They also provided a phase diagram for transient operation of bubble production in yield stress fluids. Wang et al.~\citep{wanga-2011} also studied different flow patterns such as slug, slug\textendash annular, and annular flow in a T\textendash junction microchannel using a gas\textendash carboxymethyl cellulose (CMC) system. We have recently reported Taylor bubble formation in a co\textendash flow microchannel with Newtonian and power\textendash law liquids using VOF method.\cite{sontti2017cfd} Different mass concentrations of CMC were considered as the non\textendash Newtonian liquid phase, and the results showed that in the presence of CMC, bubble length decreased, but formation frequency increased due to enhanced effective viscosity of the liquid phase. We also reported flow pattern maps in co\textendash flow microchannel based on gas and liquid phase inlet velocities for power\textendash law liquids.\cite{sontti2017b} This work aims to investigate the characteristics of Taylor bubble flow in high viscous liquids. Taylor bubble formation in non\textendash Newtonian CMC solutions is studied using finite volume method. To understand the underlying physics of Taylor bubble formation in shear thinning liquids, a Carreau\textendash Yasuda viscosity model is considered, which has not been addressed earlier in literature. For a better understanding of the non\textendash Newtonian flow field associated with Taylor bubbles, the effects of rheological properties, liquid inlet velocity, and surface tension are studied by analyzing the bubble length, velocity, shape, surrounding liquid film thickness, and velocity field distribution inside the Taylor bubble.

\section*{Numerical model}
\noindent  In general, the physical process of multiphase flow is described by a set of conservation equations like mass, momentum, and a marker function to identify the fluid-fluid interface. As discussed in the previous section, there are several interface capturing methods available with its own advantages and disadvantages. VOF model is relatively simple but accurate enough that accounts for substantial topology changes of the interface.\citep{van2006numerical, pathak2011numerical, zhang2014effects, ren2015breakup, lei2016simulation} It has also been proven that VOF method precisely tracked the interface with relatively lesser computational effort.\citep{van2006numerical} Recently, we also have successfully implemented the VOF method for droplet/bubble formation in Newtonian and non-Newtonian liquids.\citep{sontti2017cfd, ssontti2017cfd2} Accordingly, in this study, VOF method is utilized, and the following set of equations are solved for mass, momentum and volume fraction calculation.

\subsection*{Volume of Fluid (VOF) method}

In VOF approach, a single set of conservation equations is solved for immiscible fluids. The governing equations of the VOF formulation for multiphase flows are as follows:\citep{hirt-1981}

\noindent \textbf{Equation of continuity:} 

\begin{equation}
\label{eq:mass_eqn}
\frac{\partial  \rho }{\partial t}  +  \nabla .  ( \rho  \vec{ U } ) =0
\end{equation}

\noindent \textbf{Equation of motion:}

\begin{equation}
\label{eq:mom_eqn}
\frac{ \partial (\rho \vec{ U })}{ \partial t} + \nabla.( \rho \vec{ U } \vec{ U }) = - \nabla P + \nabla.\overline{\overline \tau} + \rho \vec{g}+ \vec{ F}_{SF}
\end{equation}

\begin{equation}
\label{eq:tau}
\overline{\overline \tau}= \eta \dot{ \gamma } = \eta (\nabla \vec {U} + \nabla { \vec {U} } ^{T})
\end{equation}

\noindent where $\vec{U }$, $\rho $, $\eta $, $P$, $\vec{g}$ and $\vec{ F}_{SF}$ are velocity, density, dynamic viscosity of fluid, pressure, gravitational acceleration, and surface tension force, respectively.

\noindent For a two\textendash phase system, if the phases are represented by the subscripts and the volume fraction ($C$) of the secondary phase is known, then the density and viscosity in each cell are calculated by: 
\begin{equation}
\label{eq:frac1_eqn}
\rho =  C_{2}  \rho _{2}  + (1 - C_{2})\rho _{1} 
\end{equation}
\begin{equation}
\label{eq:frac2_eqn}
\eta =  C_{2}   \eta  _{2}  + (1 - C_{2}) \eta  _{1} 
\end{equation}

\noindent \textbf{Equation of marker function:}

\noindent In absence of any mass transfer between phases, the interface between the two phases can be tracked by solving the following continuity equation (Equation  \ref{eq:vof_eqn}) for the volume fraction function.
\begin{equation}
\label{eq:vof_eqn} 
\frac{\partial    C _{q}  }{\partial t}  +  ( \vec{ U_q }  .  \nabla C_{q})  =0
\end{equation}

\noindent where $ q$ denotes either gas or liquid phase. The volume fraction for the primary phase in Equation  \ref{eq:vof_eqn} is then obtained from the following equation:
\begin{equation}
\sum  C_{q} =1
\end{equation} 

\noindent \textbf{Continuum surface tension (CSF) model:}

\noindent The continuum surface force (CSF) model\citep{brack-1992} that has been widely and successfully applied to incorporate surface tension force, is used in this work. Surface tension force ($\vec{ F}_{SF}$) is added to VOF calculation as a source term in the momentum equation. For gas\textendash liquid two\textendash phase flows, the source term in Equation \ref{eq:mom_eqn} that arises from surface tension can be represented as: 

\begin{equation}
\vec{F} _{SF} = \sigma  \kappa_{n}  \begin{bmatrix} \frac{  C _{1}  \rho _{1}+   C _{2}  \rho _{2}}{ \frac{1}{2}  ( \rho_{1}+  \rho_{2})}   \end{bmatrix} 
\end{equation}

\noindent where $\kappa_{n}$ is the radius of  curvature and  $\sigma$ is the surface tension. $\kappa_{n}$ is further defined in terms of the unit normal $ \hat{N} $ as follows \citep{fluent} :

\begin{equation}
\kappa_{N} = - \nabla  .  \hat{N}= \frac{1}{|\vec{ N}|}  \begin{bmatrix} \big(\frac{\vec{ N}}{|\vec{ N}|} . \nabla\big) |\vec{ N}| - \big( \nabla  . \vec{ N}\big) \end{bmatrix}
\end{equation}

\noindent where $\hat{N} = \frac{\vec N}{ |\vec N | } $, and $\vec N= \nabla $C$ _{q}$.

\noindent If $ \theta _{ w } $ is the contact angle at the wall, then the surface normal at the live cell next to the wall is
\begin{equation}
\hat{N}=  \hat{N} _{ w }  cos  \theta _{ w }  +  \hat{M} _{ w } sin  \theta _{ w } 
\end{equation}

\noindent where $\hat{N}$ and $\hat{M}$ are the unit vectors normal and tangential to the wall, respectively \citep{fluent}.
\subsubsection*{Constitutive equation of continuous phase}
\noindent To implement shear thinning nature of the liquid phase, Carreau\textendash Yasuda viscosity model\citep{carreau1972} has been implemented, where the effective viscosity ($\eta_{eff}$) is expressed as:
\begin{equation}
\label{eq:nnvis_eqn}
\eta_{eff}= \eta_ \infty+ \big(\eta_0- \eta_\infty \big) \Big[1+  {\big(\lambda \dot{\gamma }\big)}^{2} \Big]^{ \frac{n-1}{2}}  
\end{equation} 

\noindent where $\dot{\gamma}$ is the applied shear rate, $\eta_0$ represents the dynamic viscosity corresponding to the zero shear rate (${\dot{\gamma }} \mapsto  0$), and $\eta_ \infty$ is the viscosity at infinite shear rate (${\dot{\gamma }} \mapsto  \infty$), which was set as the solvent viscosity. The parameter $\lambda$ denotes the relaxation time, and $n$ is the flow behaviour index.

\subsection*{Computational model}
\begin{figure}[!h]
	\centering
	\includegraphics[width=1\textwidth,height=0.50\textheight,keepaspectratio]{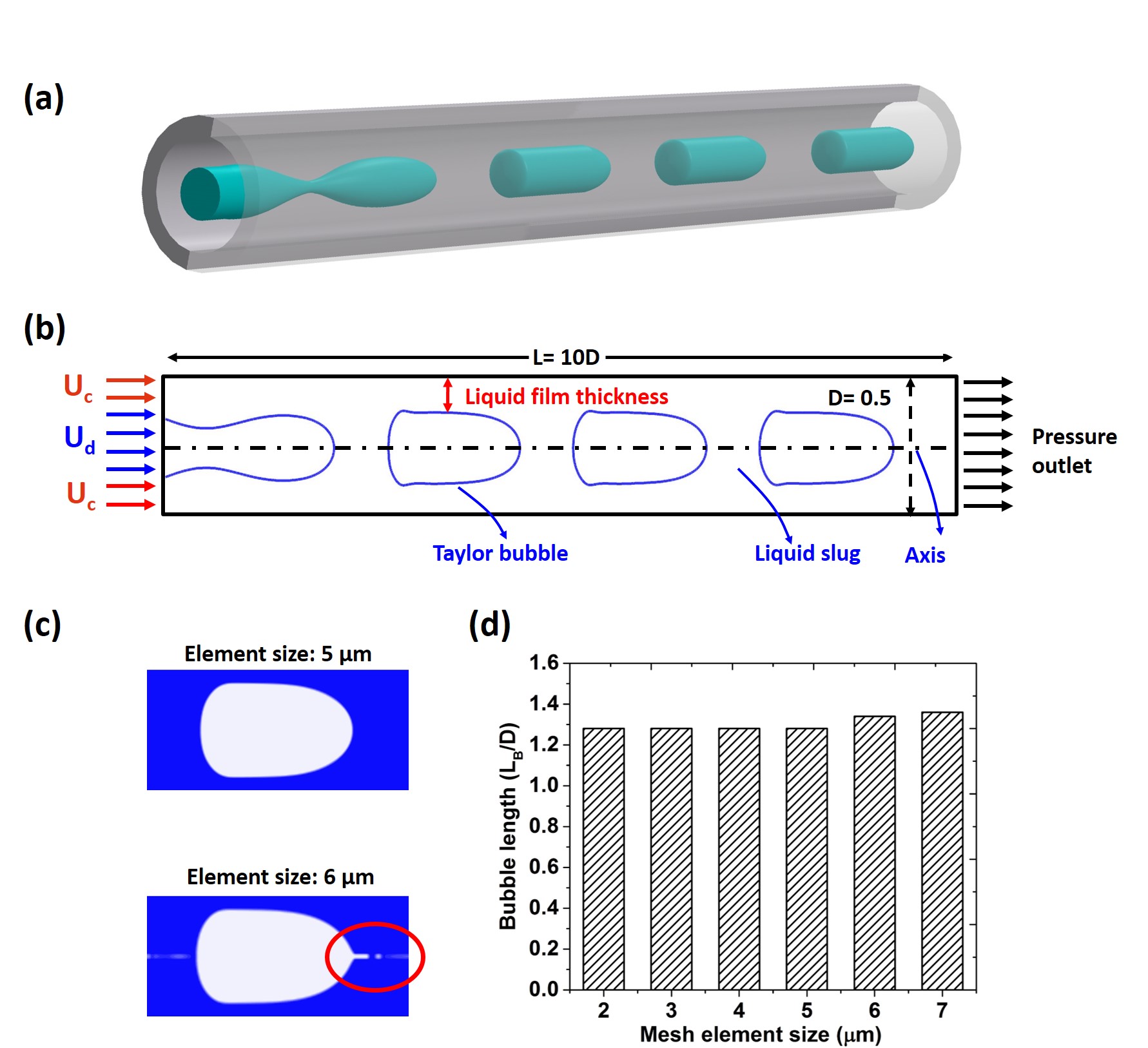}
	\caption{\label{fig:M1} (a) 3D schematic of the Taylor bubble formation in circular co\textendash flow microchannel, (b) 2D representation of the computational domain with imposed boundary conditions, (c) interface tracking comparison between two different mesh element sizes (5~$\mu m$ vs. 6~$\mu m$), highlighting elongated thread in coarse mesh (6~$\mu m$) and, (d) grid independence study of the bubble length for air\textendash CMC\textendash1.0\% system for $U_G=0.5~m/s$, $U_L=0.5~m/s$.}
\end{figure}
\noindent A 3D schematic of Taylor bubble formation in circular co\textendash flow microchannel is illustrated in Figure~\ref{fig:M1}a, where the gas bubbles are separated by thin liquid film adjacent to the wall. In this work, a circular microchannel of diameter (D) 0.5 mm, and a length of 10~D is considered. The selection of $L/D=10$ is based on the modelling guideline and practice followed by previous researchers.\citep{gupta-2009, gupta2010cfd, asadolahi2011cfd, sontti2017cfd} Our preliminary simulations also indicate that higher ratio than $L/D=10$ does not induce any significant influence on bubble length, velocity, and surrounding film thickness for the considered ranges of operating conditions. The computational domain is taken into account as a two\textendash dimensional axisymmetric geometry (Figure~\ref{fig:M1}b). Transient simulations are carried out in a finite volume method based solver, ANSYS Fluent 17.0, to solve aforementioned partial differential equations. The pressure\textendash velocity coupling is approximated by fractional step method (FSM) using first\textendash order implicit non\textendash iterative time advancement (NITA) scheme.\citep{fluent} Quadratic upstream interpolation for convective kinetics (QUICK) and geo\textendash reconstruct schemes are used for the momentum and volume fraction equation discretization, respectively. Variable time step and fixed Courant number (Co = 0.25) are used for solving momentum and pressure equations. At the liquid and gas inlets, constant velocity is imposed, and the pressure outlet boundary is set at the microchannel outlet. No\textendash slip condition is applied to the impermeable wall. Additional details on the model implementation can be found in our earlier work.\citep{sontti2017cfd} At first, grid independence study was performed with different mesh element sizes, varying from $7~\mu m$ to $2~\mu m$. Smooth gas-liquid interface was not captured in simulations with coarser grids ($6~\mu m$ and $7~\mu m$), and Figure.~\ref{fig:M1}c illustrates the appearance of elongated thread at the nose of Taylor bubble in such cases. Figure.~\ref{fig:M1}d demonstrates the influence of mesh element size on the bubble length, which did not change with mesh sizes below $5~\mu m$ under identical operating conditions. Therefore, optimum mesh element size is taken to be 5 $\mu m$. The results presented henceforth are based on element size of 5 $\mu m$ in the core region. To capture the sharp interface and the liquid film thickness around the Taylor bubble, mesh elements near the wall are further refined, as per the guidelines of Gupta et al.\citep{gupta-2009} Subsequently, the predicted liquid film thickness in the near wall region are analyzed.  

\section*{Results and discussion}
\begin{figure}[!h]
	\centering
	\includegraphics[width=1\textwidth]{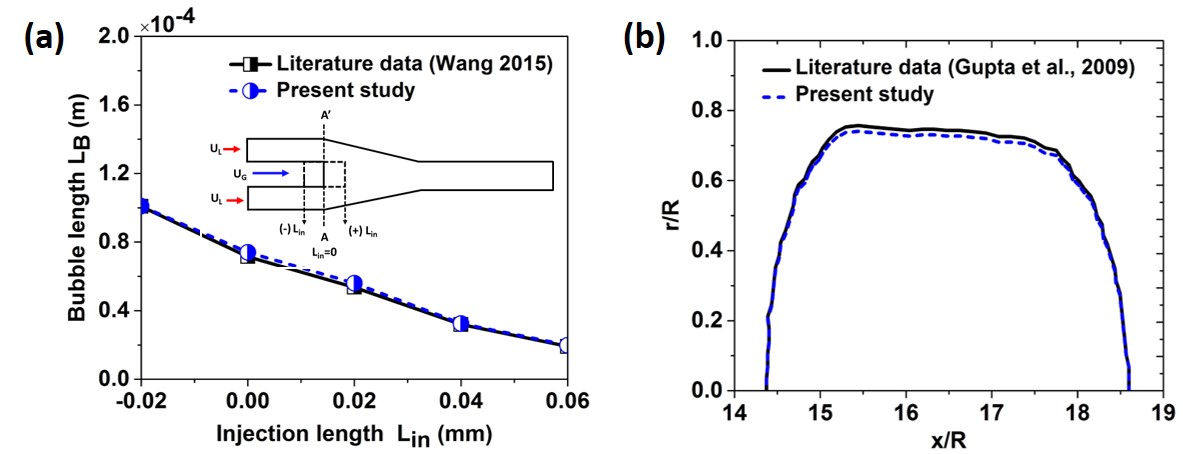}
	\caption{\label{fig:M2} (a) Comparison of Taylor bubble length for different injection length  ($L_{in}$) for $\eta_w$= 1.003$\times10^{-3}$ kg/m.s, $Q_{G}= 0.47$ $\mu L/s$, and $Q_{L}= 2.01$ $\mu L/s$ with the results of Wang~\citep{wangg-2015}, and (b) comparison of Taylor bubble shape with Gupta et al.~\citep{gupta-2009} for air\textendash water system with $U_G=0.5~m/s$, $U_L=0.5~m/s$.}
\end{figure}

\noindent The developed CFD model is initially validated with the literature data of Wang~\citep{wangg-2015} and Gupta et al.~\citep{gupta-2009} for air\textendash water two\textendash phase flow systems in co\textendash flow microchannels. For validation with Wang~\citep{wangg-2015}, a tapered co\textendash flow configuration is simulated. The model predictions of bubble length are found to be in excellent agreement with literature data, as shown in Figure~\ref{fig:M2}a.  Furthermore, the Taylor bubble shape in a straight circular microchannel is compared with the results of Gupta et al.~\citep{gupta-2009} to reinforce credibility of the developed model. Notably, for Ca=0.006, the bubble velocity (0.55 m/s) and shape are found to be identical with literature data, as shown in Figure~\ref{fig:M2}b. These two validations with independent sources advocate the efficacy of this model, and is extended for the present study. 

\subsection*{Effect of Carboxylmethyl cellulose (CMC) concentration }
\noindent In this section, the effect of CMC concentration on Taylor bubble formation has been systematically explored. To understand the rheological properties using Carreau\textendash Yasuda viscosity model, we have considered four different mass concentrations of CMC in water. Physical properties of different CMC solutions are obtained from the experimental data of Sousa et al.~\citep{sousa2005}, and are summarized in Table~\ref{my-label}.
 
\begin{table}[h]
	\centering
	\caption{ Rheological properties of Carreau\textendash Yasuda model parameters for different CMC solutions\cite{sousa2005}.}
	\vspace{0.5cm}
	\label{my-label}
	\begin{tabular}{lllllll}
		\hline
		CMC (wt\%) & $\rho$(kg/$m^{3}$) & $\eta_0$ (Pa.s)  & $\eta_ \infty$ (Pa.s) & $\lambda$ (s) & $n$  (\textendash)&  $\sigma$ (N/m) \\ \hline
		0.1& 996.28&0.0091               & 0.001               & 0.0214            &0.87& 0.072   \\ 
		0.4	& 995.39&0.1102               &  0.001              & 0.1099            &0.67 & 0.072  \\ 
		0.6	&994.13&  0.3602              &  0.001              &  0.1828           &0.57& 0.072   \\
		1.0	&993.61& 2.9899               &0.001                &0.3653             & 0.40 & 0.072  \\ \hline
	\end{tabular}
\end{table}

\noindent It can be noted from Table~\ref{my-label} that the surface tension value is identical for all solutions, which is similar to that of water. To understand the viscous effects of non\textendash Newtonian liquids, the effective viscosity ($\eta_{eff}$) is calculated based on the rheological properties (Equation~\ref{eq:effective_eqn}\citep{shahsavari2015mobility}), and the influence of CMC concentration on Taylor bubble formation is explained based on $\eta_{eff}$ of the continuous phase.
\begin{equation}
\label{eq:effective_eqn}
\eta_{eff}=(\eta_0- \eta_\infty)\lambda^{n-1} \left(\frac{3n+1}{4n}\right)^{n}\left(\frac{8U_{L}}{D}\right)^{n-1}
\end{equation}
\noindent where $\eta_{0}$, $U_L$, $D$, and $n$ are consistency index, liquid velocity, diameter of the channel, and power\textendash law index, respectively. 

\begin{figure}[H]
	\centering
	\includegraphics[width=0.9\textwidth]{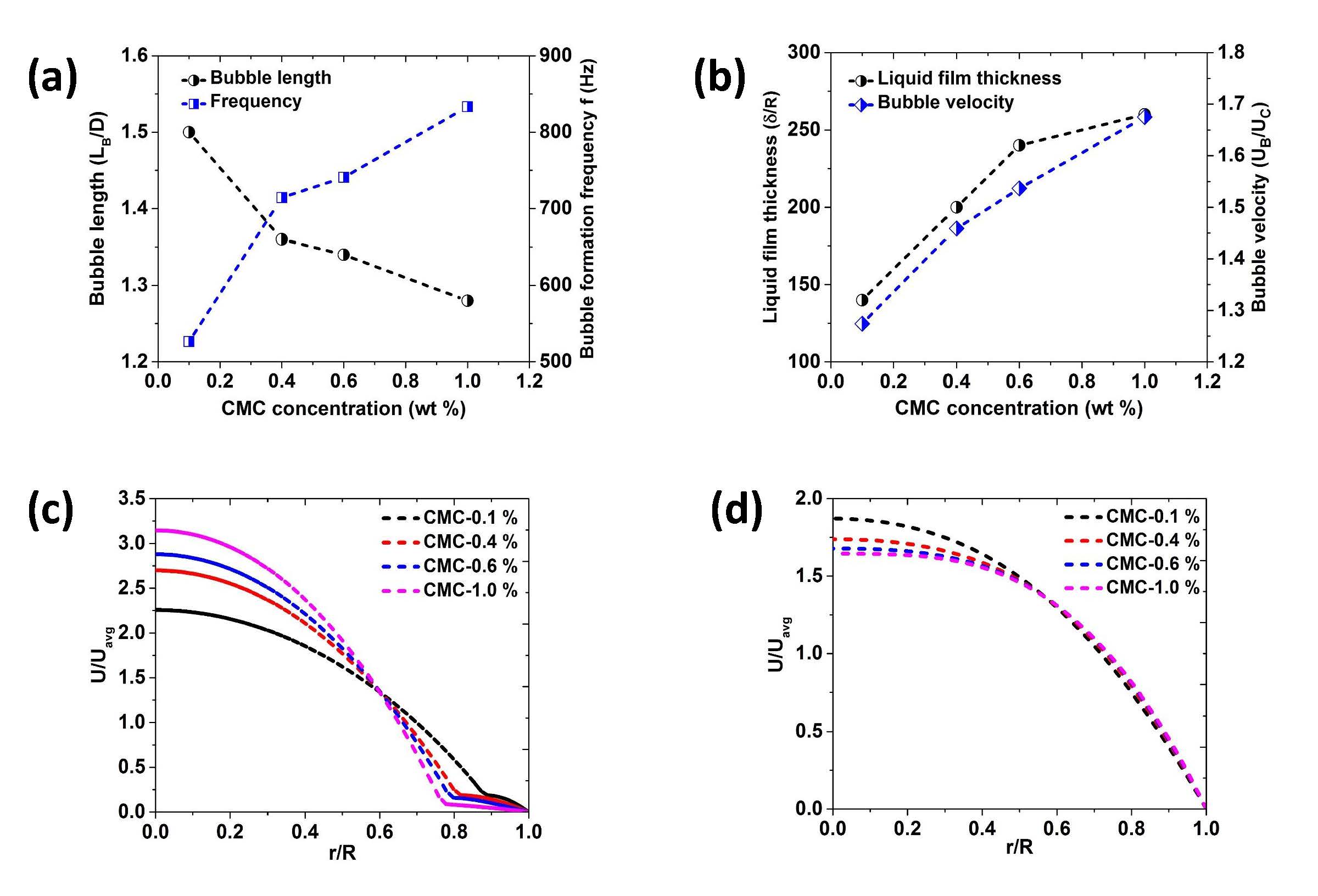}
	\caption{\label{fig:CMC2}Effect of CMC concentration on (a) non\textendash dimensional bubble length and formation frequency, (b) surrounding liquid film thickness, (c) velocity profile in the middle of a Taylor bubble, and (d) velocity profile in the middle of a liquid slug at $U_{L}$ = 0.5 m/s and $U_{G}$ = 0.5 m/s.}
\end{figure}

\noindent Figure~\ref{fig:CMC2}a shows that on increasing CMC concentration, Taylor bubble length decreases, and the formation frequency increases due to increasing effective viscosity of the liquid phase. A thin liquid film around the Taylor bubble is precisely captured, which is measured from wall to the middle of the bubble. Surrounding liquid film thickness is observed to increase, as shown in Figure~\ref{fig:CMC2}b. Consequently, Taylor bubble velocity increases due to increase in liquid film thickness and change in the bubble shape/nose curvature, which decreases with increasing CMC concentration. A detailed insight on velocity fields inside the Taylor bubble and liquid slug are necessary to understand the heat and mass transfer performance. To realize the viscous effects, non\textendash dimensional velocity profiles are analyzed in the middle of the Taylor bubble and liquid slug, and are illustrated in Figure~\ref{fig:CMC2}c and Figure~\ref{fig:CMC2}d, respectively. 

\begin{figure}[h]
	\centering
	\includegraphics[width=0.65\textwidth]{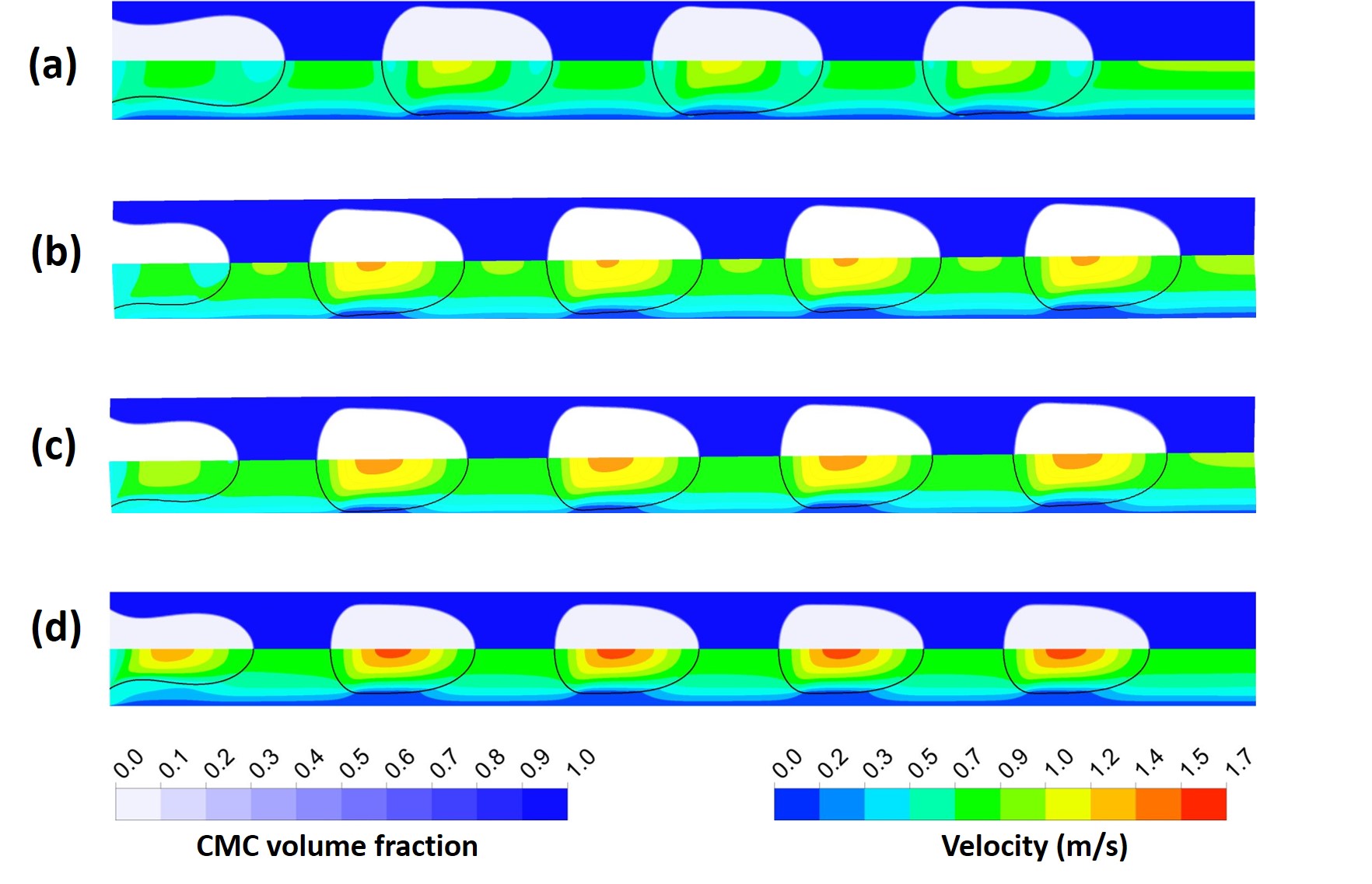}
	\caption{\label{fig:CMC3} Effect of CMC concentration on velocity distribution (upper halves are volume fraction and lower halves are velocity field) for (a) CMC\textendash 0.1~\%, (b)CMC\textendash 0.4~\%, (c) CMC\textendash 0.6~\%, and (d) CMC\textendash 1.0~\% at $U_{L}$ = 0.5 m/s and $U_{G}$ = 0.5 m/s.}
\end{figure}

\begin{figure}[!h]
	\centering
	\includegraphics[width=0.65\textwidth,keepaspectratio]{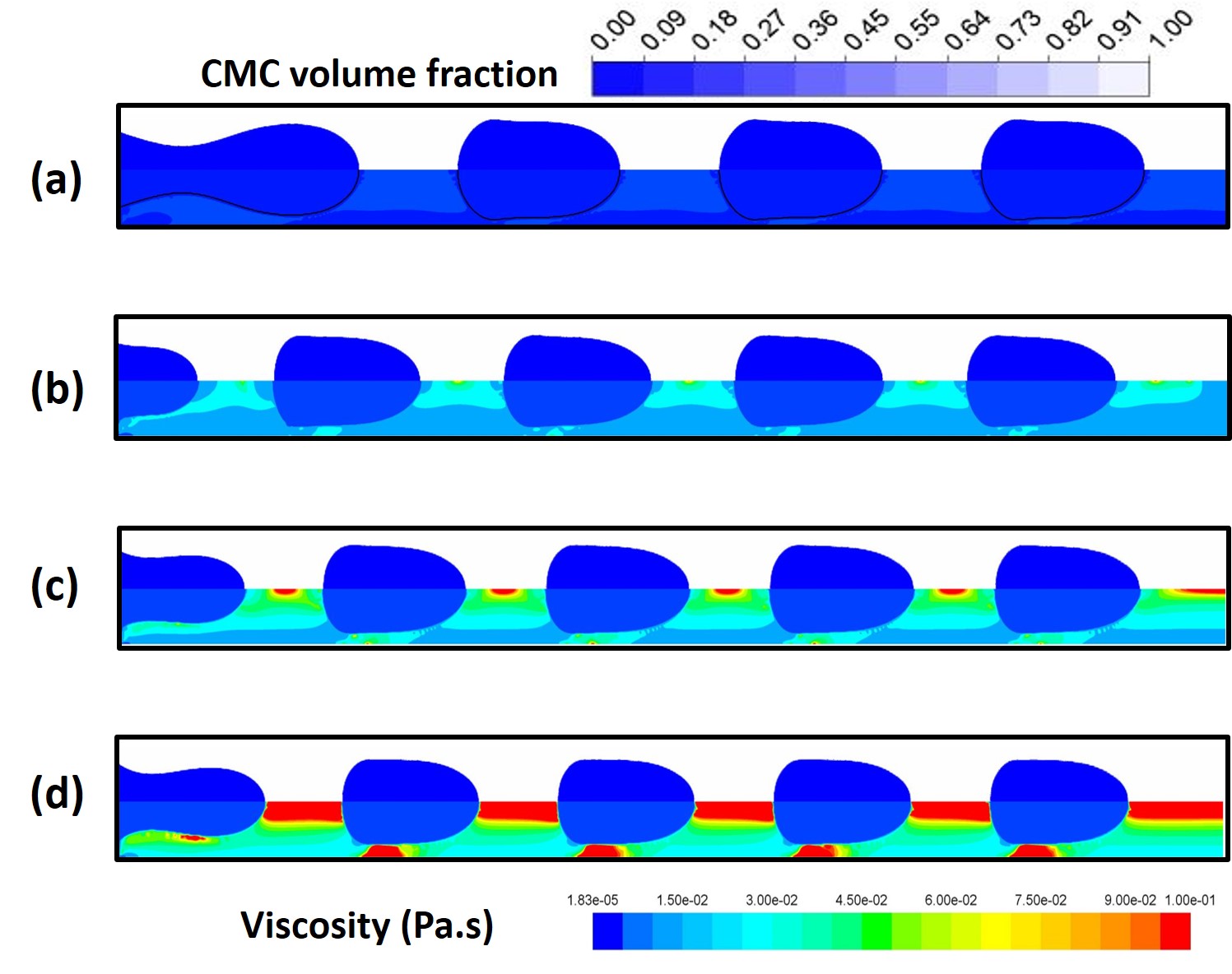}
	\caption{\label{fig:VIS} Effect of CMC concentration on non-homogeneous viscosity distribution (upper halves are volume fraction and lower halves are viscosity distribution) for (a) CMC\textendash 0.1~\%, (b)CMC\textendash 0.4~\%, (c) CMC\textendash 0.6~\%, and (d) CMC\textendash 1.0~\% at $U_{L}$ = 0.5 m/s and $U_{G}$ = 0.5 m/s.}
\end{figure}

The velocity inside a Taylor bubble is found to increase with increasing CMC concentration, as shown in Figure~\ref{fig:CMC2}c. Liquid film thickness surrounding the bubble is characterised by a discontinuous velocity field at the gas-liquid interface. Figure~\ref{fig:CMC2}d depicts the effect of CMC concentration on velocity profiles of the continuous liquid phase. In the case of lower concentrations, the nearly parabolic profile is observed in the liquid slug. However, a relatively flatter profile is depicted at the higher concentration, which is the typical characteristic of shear thinning liquids. Figure~\ref{fig:CMC3} illustrates that with increasing CMC concentration, velocity field inside the Taylor bubble significantly increases, which corresponds to the quantitative analyses in Figure~\ref{fig:CMC2}c.
Subsequently, the non-homogeneous viscosity distribution in the microchannel is analyzed for considered CMC solutions. From Table~\ref{my-label} and Equation~\ref{eq:effective_eqn}, it is apparent that the effective viscosity increases with increasing concentration of CMC. Figure~\ref{fig:VIS} describes viscosity distribution along with volume fraction for various CMC solutions, and as expected, it is closely related to velocity field around the bubble, as shown in Figure~\ref{fig:CMC3}. With increasing the CMC concentration, the maximum magnitude of effective viscosity around the Taylor bubble significantly increases, as shown in Figure~\ref{fig:VIS}d.  

\begin{figure}[!h]
	\centering
	\includegraphics[width=0.45\textwidth,height=0.50\textheight,keepaspectratio]{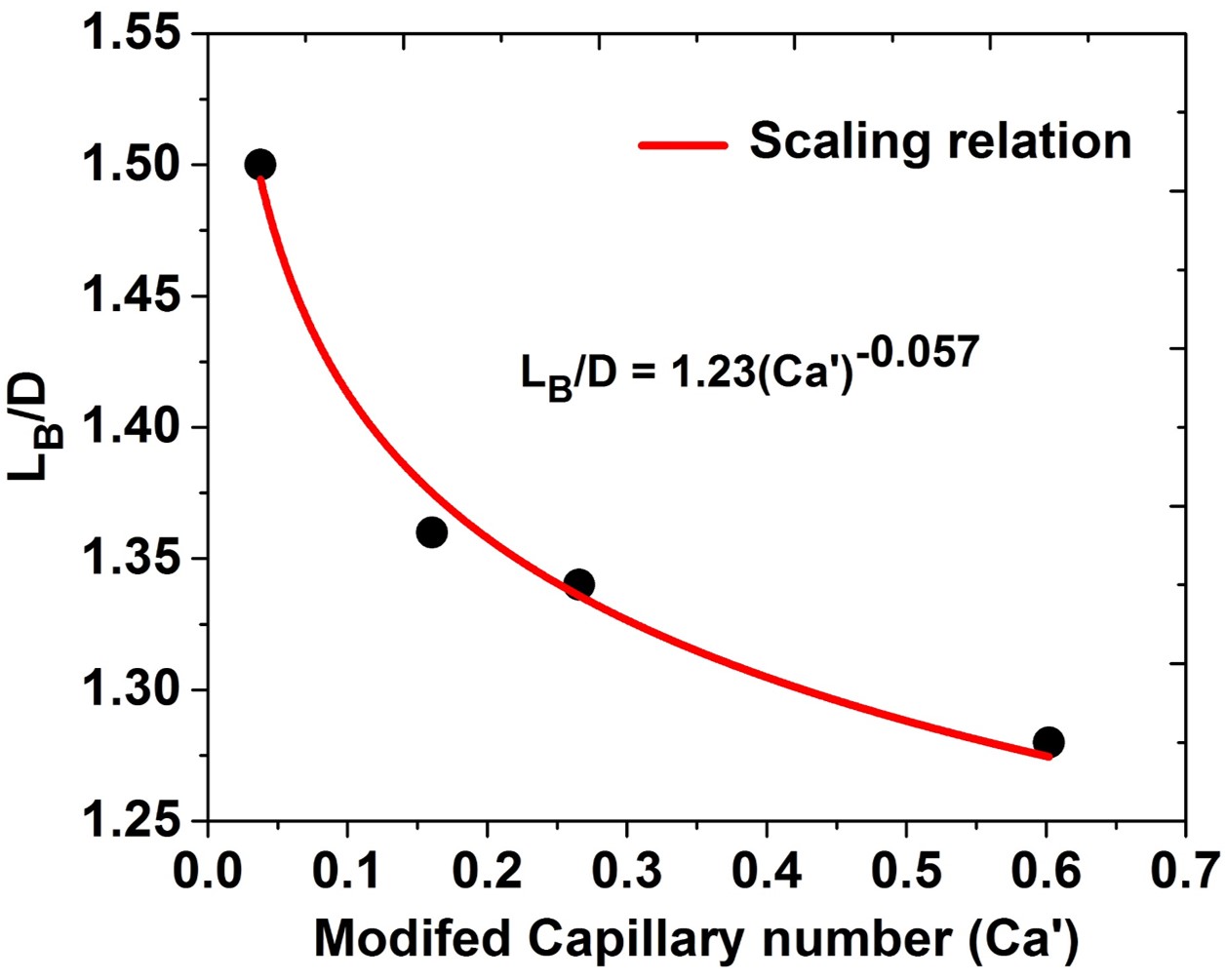}
	\caption{\label{fig:SSA} Non\textendash dimensional bubble length as a function of modified Capillary number for different CMC solutions at $U_{L}$ = 0.5 m/s and $U_{G}$ = 0.5 m/s.}
\end{figure}

\noindent Additionally, a scaling relation is derived to determine the non\textendash dimensional bubble length as a function of modified Capillary number \citep{shahsavari2015mobility} ($Ca^{'}=\big(\eta_0- \eta_\infty \big)\lambda^{n-1}U^nD^{(1-n)}/\sigma$) for the range of CMC concentration from 0.01\% to 1.0\% (i.e., $Ca^{'}=0.037-0.602$) in Figure~\ref{fig:SSA}, where gas and liquid inlet velocities are kept constant at 0.5 m/s). The proposed scaling law relation $L_B/D=1.23(Ca^{'})^{-0.057}$ for different CMC solutions shows a maximum deviation of 1.2~\%, and is in line with other non\textendash Newtonian studies \cite{fu-2011,ssontti2017cfd2}, but with different pre-factor and exponent. 

\subsection*{Effect of liquid phase velocity}
\begin{figure}[!h]
	\centering
	\includegraphics[width=0.9\textwidth]{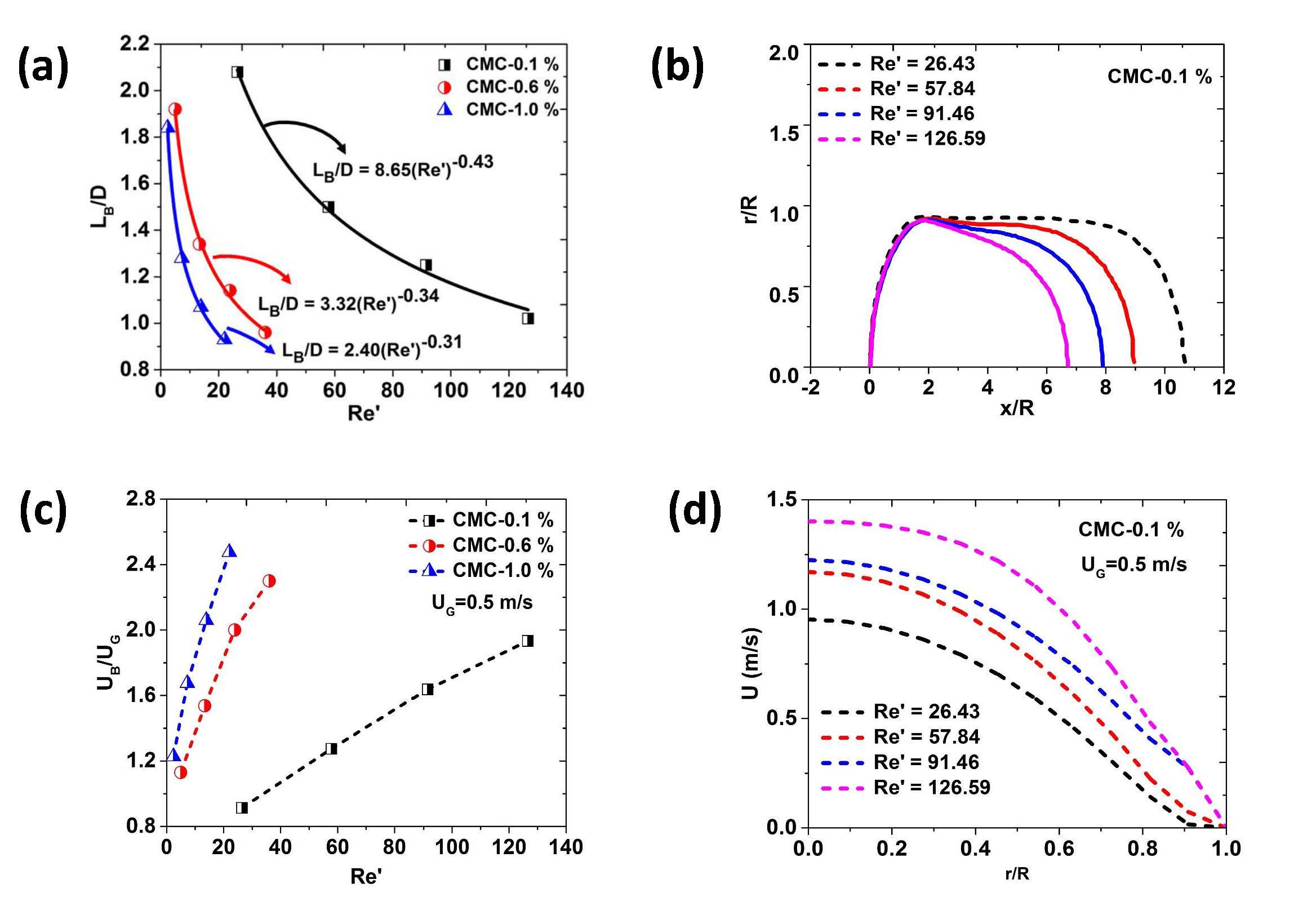}
	\caption{\label{fig:UL1} Effect of liquid phase inlet velocity on (a) non\textendash dimensional bubble length, (b) Taylor bubble shape for CMC\textendash0.1~\%, (c) bubble velocity, and (d) velocity profiles inside Taylor bubble at $U_{G}$=0.5 m/s for CMC\textendash0.1~\%.}
\end{figure}

\noindent In this section, effect of liquid phase velocity of three different CMC concentration solutions on bubble length, velocity, shape and velocity field inside the Taylor bubble are systematically studied. It can be observed from Figure~\ref{fig:UL1}a, that bubble length decreases with increasing liquid phase inlet velocity due to increase in inertial force on gas phase. Consequently, scaling relations are proposed for non-dimensional bubble length as a function of modified Reynolds number ($Re^{'}=\rho U_{L}D/\eta_{eff}$), and are mentioned in Figure~\ref{fig:UL1}a. It can also be realized that the exponent and pre-factor of the proposed power-law relation change systematically with varying CMC concentration. It is worth mentioning that the proposed relations predict with high degree of confidence, when the gas phase inlet velocity is kept constant at 0.5~m/s, as the $Re^{'}$ is calculated based on the liquid phase properties. At lower liquid velocity, a considerable change in the bubble size is observed for three different CMC liquids. Effect of liquid velocity on Taylor bubble shape is also analyzed for CMC\textendash 0.1\%. From Figure~\ref{fig:UL1}b, it can be seen that bubble nose shape significantly changes with increasing liquid phase inlet velocity. Consequently, bubble velocity increases due this change in bubble shape and surrounding liquid film thickness (Figure~\ref{fig:UL1}c). Velocity inside the Taylor bubble also increases with enhanced liquid phase inlet velocity, as depicted in Figure~\ref{fig:UL1}d. This phenomenon is also attributed to the consequence of bubble shape and liquid film thickness variation. For CMC\textendash 0.1\%, at lower liquid phase velocity, elongated bubble and lower velocity field are observed inside the Taylor bubble.

\begin{figure}[H]
	\centering
	\includegraphics[width=0.9\textwidth]{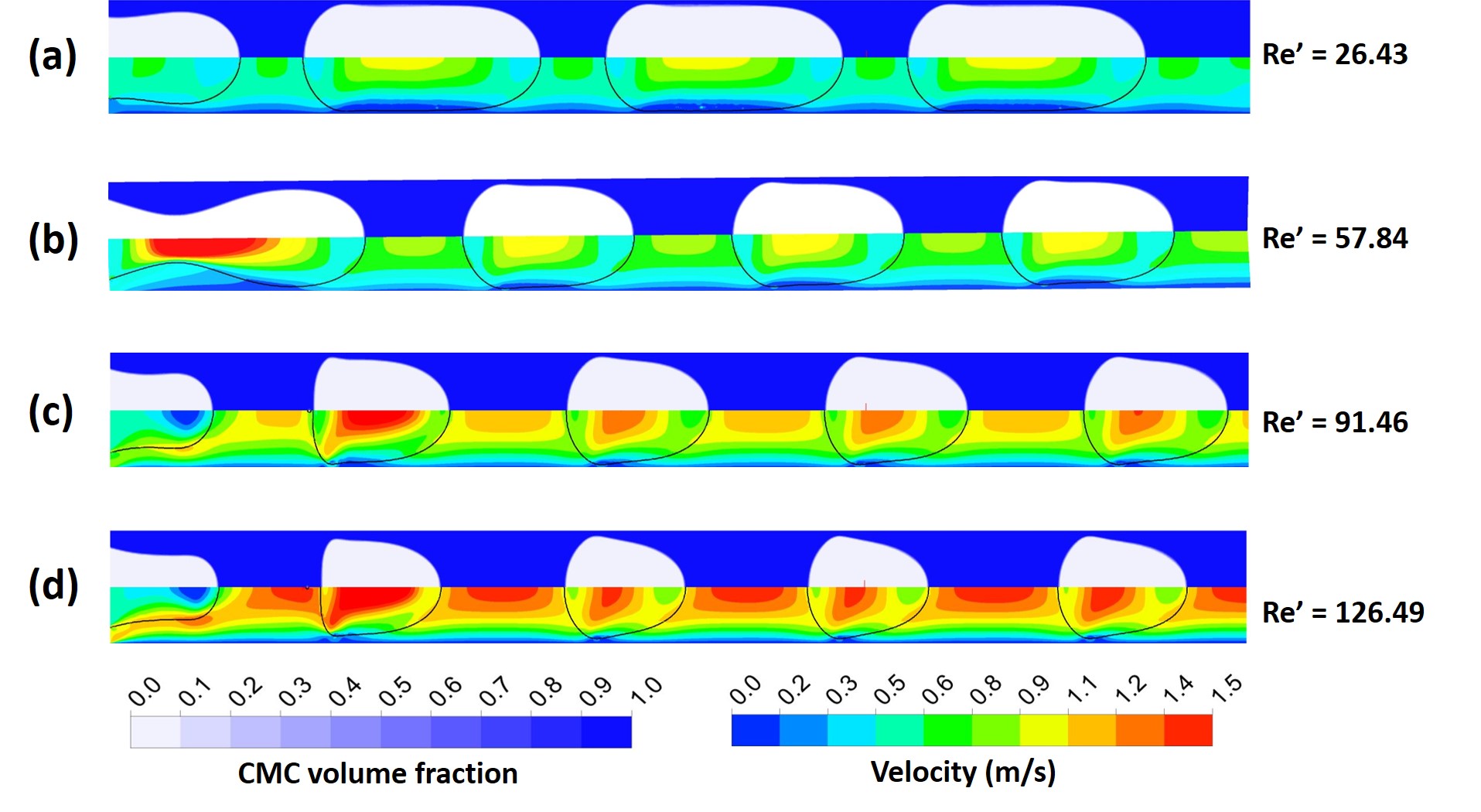}
	\caption{\label{fig:UL2} Effect of liquid flow rate on velocity distribution (upper halves are volume fraction and lower halves are velocity field) in CMC\textendash 0.1~\% for fixed $U_{G}$=0.5 m/s and at $Re^{'}$ = (a) 26.43, (b) 57.84, (c) 91.46, and (d) 126.49. }
\end{figure}

\noindent 
At a fixed gas inlet velocity, the radius of bubble nose curvature becomes smaller, and the rear cap turns flatter with increasing liquid inlet velocity, as shown in Figure~\ref{fig:UL2}. It can also be observed from Figure~\ref{fig:UL2} that velocity magnitude in the slug region, and in the middle of a bubble increases systematically with increasing liquid phase inlet velocity.  

\subsection*{Flow regimes maps}

\begin{figure}[!h]
	\centering
	\includegraphics[width=\textwidth]{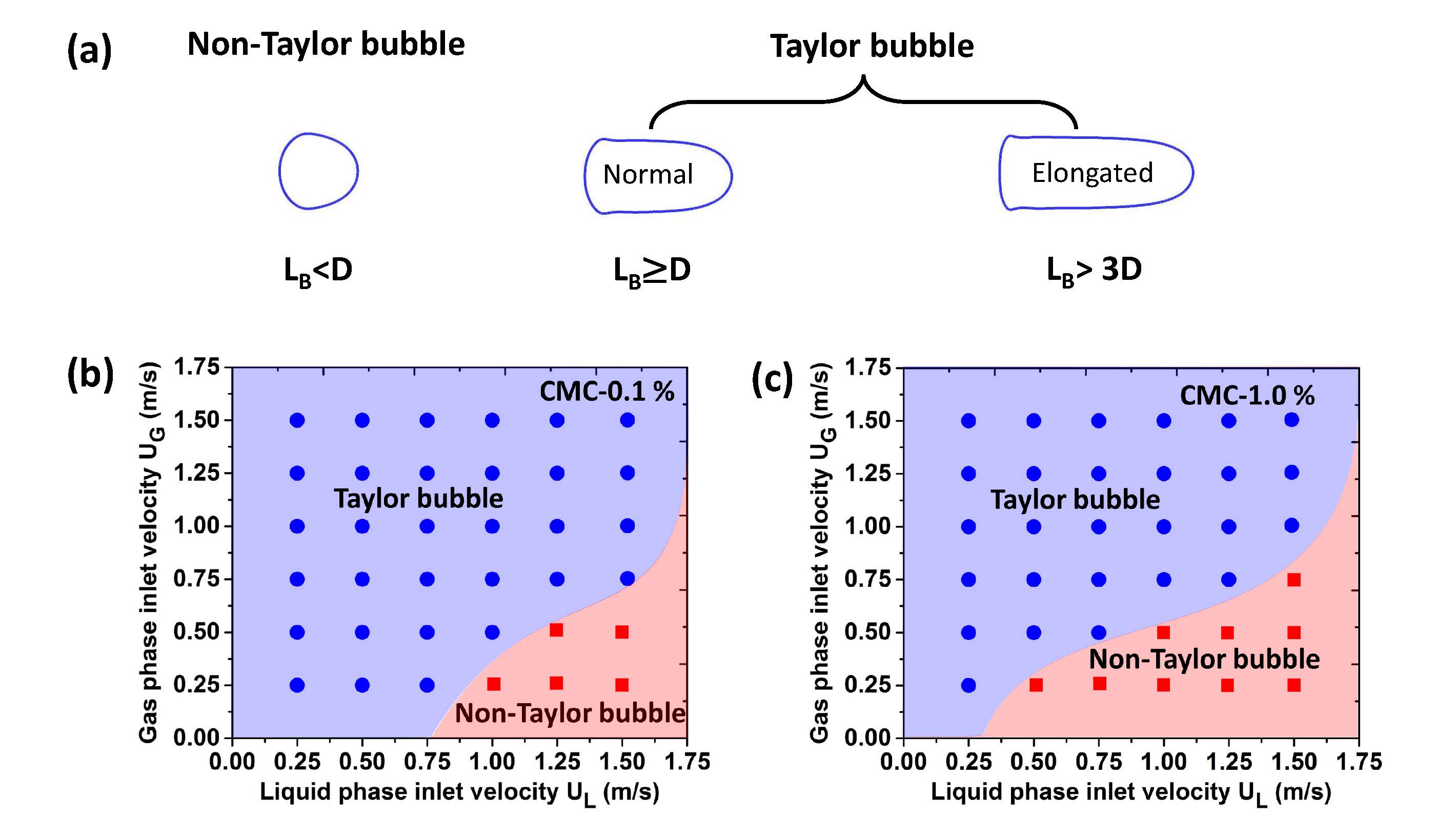}
	\caption{\label{fig:RR1} (a) Different shapes of Taylor bubbles, and flow regime map for (b) CMC\textendash0.1~\%, and (c) CMC\textendash1.0~\%.}
\end{figure}

\noindent In two\textendash phase flow systems, knowledge of flow patterns is essential for understanding the behaviour of gas\textendash liquid flows at a given operating condition. It can be recognized from the literature that reported flow regime maps in two\textendash phase flows are typically valid for only Newtonian systems. Therefore, it is necessary to develop flow regime maps for gas\textendash non\textendash Newtonian liquid systems. Here, flow regimes are broadly categorized based on the bubble length into two main types such as, non\textendash Taylor bubble, where the bubble length is smaller than the capillary diameter ($L_{B}$\textless D), and the Taylor bubble ($L_{B}$\textgreater D), as illustrated in Figure~\ref{fig:RR1}a. Flow regime maps for two different CMC solutions (CMC\textendash 0.1 and CMC\textendash 1.0~\%) under various gas and liquid velocities are portrayed in Figure~\ref{fig:RR1}b and Figure~\ref{fig:RR1}c, where the inlet velocities of the dispersed and continuous phases are used as the ordinate and abscissa, respectively. Figure~\ref{fig:RR1}b shows that for lower CMC concentration (CMC\textendash 0.1~\%), Taylor bubble regime occupies a larger area in the flow regime map and non\textendash Taylor bubbles are mostly observed at higher liquid and lower gas\textendash inlet velocities. However, with increasing concentration (CMC\textendash 0.1~\%) the appearance of non-Taylor bubble can be observed even at lower liquid inlet velocities, as shown in Figure~\ref{fig:RR1}c. In all cases, elongated Taylor bubbles are determined at higher gas inlet velocity, which is also labelled as Taylor bubble regime in this study (Figure~\ref{fig:RR1}a). It can be noted from Figure~\ref{fig:RR1}c, that with increasing CMC concentration, non\textendash Taylor bubble regime expands as compared to lower concentration of CMC solution (Figure~\ref{fig:RR1}b).         

\subsection*{Effect of surface tension}
\begin{figure}[!h]
	\centering
	\includegraphics[width=1\textwidth]{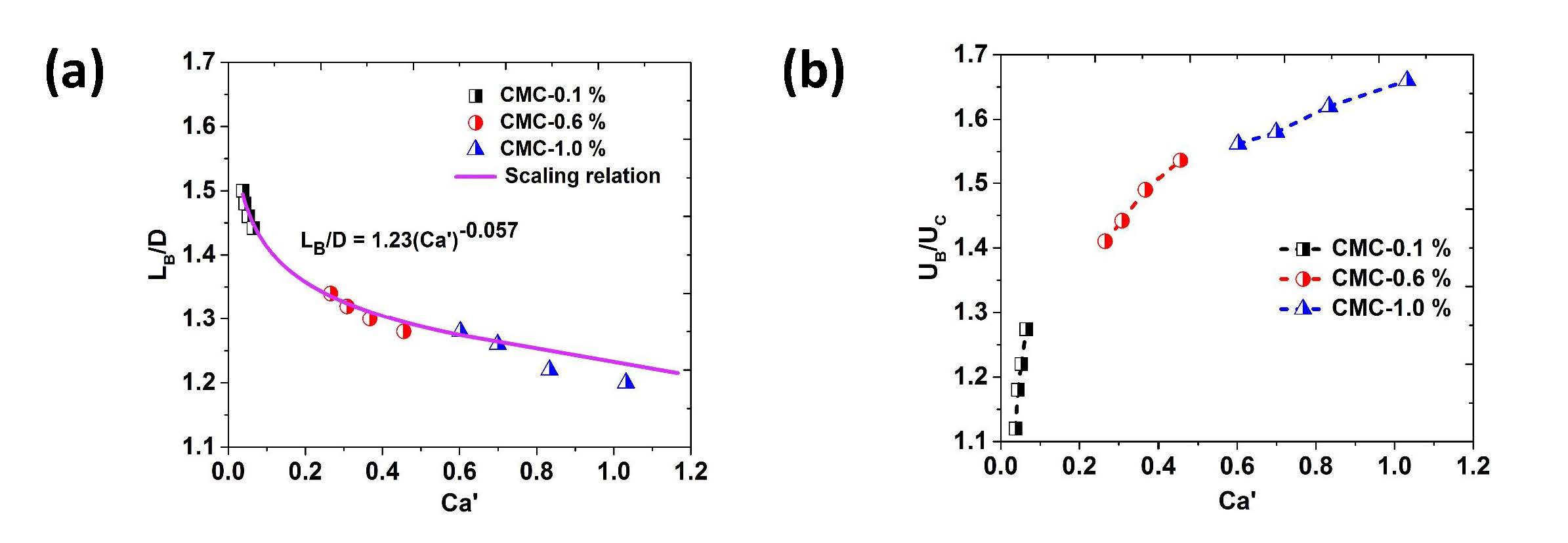}
	\caption{\label{fig:Ca1} Effect of surface tension on (a) non\textendash dimensional bubble length, and (b) bubble velocity at $U_{L}$ = 0.5 m/s, and $U_{G}$ = 0.5 m/s.}
\end{figure}

\noindent Taylor bubble length and its formation strongly depend on the surface tension and viscous forces. Several researchers have reported the influence of surface tension on air\textendash water system using different concentrations of sodium dodecyl sulfate (SDS). In this study, we systematically investigate the effect of surface tension for shear thinning liquids by altering it from 0.072\textendash 0.042 N/m. Generally, at lower surface tension, weak interfacial forces will act on both phases. The non\textendash dimensional bubble length ($L_{B}/D$) in all CMC solutions are plotted as a function of modified Capillary number ($Ca^{'}$) in Figure~\ref{fig:Ca1}a, which shows that with decreasing surface tension (i.e., increasing $Ca^{'}$), bubble length decreases in all CMC solutions. This can be attributed to the fact that at lower surface tension (higher $Ca^{'}$), the growth of dispersed phase in the microchannel is hindered by higher shear force, which in turn results into smaller bubble length. It is worth noting that the scaling law proposed in Figure~\ref{fig:SSA} corroborates to predicted values here satisfactorily with a maximum deviation of 1.2\% in range of $Ca^{'}=0.037-0.602$ for which it was developed. Interestingly, it is also in well agreement with a maximum deviation of 6.8\%, when it is extended for higher $Ca^{'} = 1.03$. In line with the discussion in previous section, bubble velocity, as depicted in Figure~\ref{fig:Ca1}b, is found to increase in all CMC solutions due to the increase in liquid film thickness and alteration in bubble shape.

\section*{Conclusions}
\noindent We have demonstrated characteristics of Taylor bubble formation in co\textendash flow microchannels by considering carboxymethyl cellulose (CMC) as a liquid phase. A CFD model based on VOF method is developed that helps in understanding the behaviour of Taylor bubble flow in Carreau\textendash Yasuda shear thinning liquids. Systematic investigations are carried out to realize the CMC concentration, liquid phase inlet velocity, and surface tension on Taylor bubble length, velocity, shape, liquid film thickness, and velocity fields inside Taylor bubble, as well as in liquid slug. On increasing the CMC concentration and liquid phase inlet velocity, Taylor bubble length was found to decrease due to increase in effective viscosity and inertial force, respectively. However, Taylor bubble velocity, liquid film thickness, and formation frequency increase with increasing CMC concentration and liquid phase inlet velocity. In all the cases, liquid film thickness between the bubble and channel wall is precisely captured to understand its effect on the bubble characteristics. Influence of CMC concentration and liquid phase inlet on velocity distribution inside the Taylor bubble and liquid slug are also presented. Three different types of bubble shapes are identified, and the flow regime maps for Carreau\textendash Yasuda liquids are developed for the first time, based on gas\textendash liquid inlet velocities. On decreasing surface tension, the Taylor bubble length was observed to decrease but the velocity increased. Scaling laws are proposed to determine the bubble length based on the modified Capillary number, and Reynolds number that take into consideration of continuous phase rheological properties, and flow rate, respectively. These findings are expected to serve as a basis for further experimental/numerical investigations with non\textendash Newtonian liquids that may contribute in the design of microfluidic devices.

\section*{Acknowledgement}
This work is supported by Science \& Engineering Research Board, Department of Science and Technology, Government of India.

\section*{Nomenclature}

$Ca^{'}$ = modified Capillary number ($(\eta_0- \eta_\infty )\lambda^{n-1}U^nD^{(1-n)}/\sigma$) \\
$Re^{'}$ = modified Reynolds number ($=\rho U_{L}D/\eta_{eff}$)\\
$D$ = diameter of the channel (m)\\
$U$  =  velocity (m/s) \\
$L$  = length (m) \\
$\hat{N}$	= unit normal vector\\
$P$ = pressure (Pa)\\
$C$= volume fraction

\noindent \textit{Greek symbol}\\
$\dot{\gamma } $ = shear rate (1/s)\\
$\delta$ = liquid film thickness (m) \\ 
$\eta$ = dynamic viscosity (Pa.s)\\
$\rho$  = density (kg/$m^{3}$)\\
$\sigma$  = surface tension (N/m)\\ 
$\overline{\overline \tau}$ = shear stress (Pa)

\noindent \textit{Subscripts}\\
$B$ = bubble \\
$G$ = gas\\
$L$ = liquid\\
\textit{eff} = effective

\section*{References}
\bibliography{sekhar}

\end{document}